\documentclass[11pt]{article}
\usepackage{authblk}
\usepackage{hyperref}
\usepackage{amsfonts}
\usepackage{mathtools}
\usepackage{amsmath}
\usepackage[dvipsnames]{xcolor}
\usepackage{verbatim}
\usepackage{hyperref}
\usepackage{cite}\usepackage[margin=2.5cm]{geometry}

\def\be{\begin{equation}}
\def\ee{\end{equation}}
\def\ba{\begin{eqnarray}}
\def\ea{\end{eqnarray}}

\begin{document}

\title{
Discovery of an insulating ferromagnetic phase of 
electrons in two dimensions}
\author{Kyung-Su Kim$^*$ \& Steven A. Kivelson$^{\dagger}$}
\affil{Department of Physics, Stanford University, Stanford, CA 93405} 
\affil{*kyungsu@stanford.edu,   ${}^\dagger$kivelson@stanford.edu}

\date{}
\maketitle

The two dimensional electron gas (2DEG) realized in semiconductor hetero-structures has  been the focus of fruitful study for decades.  It is, in many ways, the paradigmatic system in the field of highly correlated electrons.
The list of discoveries that have emerged from such studies and have opened new fields of physics is extraordinary, including discoveries related to the integer and fractional quantum Hall effects, weak localization, metal-insulator transitions, Wigner crystalization, mesoscopic quuantum transport phenomena, etc.  
Now, a set of recent studies\cite{Hossain2020ferromagnetism, hossain2020valley} on ultra-clean modulation-doped AlAs quantum wells have uncovered a new set of ordering transitions associated with the onset of ferrromagnetism and electron nematicity (or orbital ordering).

The 2DEG in the current generation of ``ultra-clean'' AlAs quantum wells have been gate-tuned over a range of electron density from $n=1.1 \times 10^{10}\textrm{cm}^{-2}$ to $n=2 \times 10^{11}\textrm{cm}^{-2}$.
The electrons occupy two valleys - one located about the $(0,\pi)$ and the other about the $(\pi,0)$ point in the 2D Brillouin zone - so in addition to two spin-polarizations, 
the electrons carry a valley ``pseudo-spin'' index. 
In the absence of shear strain, the 2DEG has a discrete $C_4$ rotational symmetry that interchanges the valleys.  Older studies of the 2DEG in Si MOSFETs and modulation-doped GaAs quantum wells have  explored the same general range of correlation strengths;  while there is considerable overlap in results -- for instance concerning the existence and character of a metal-insulator transition (MIT) -- a number of aspects 
have been seen here for the first time. 
Since each realization of the 2DEG differs in some details -- e.g. the character and strength of the disorder, the geometry of the device,  the existence of a valley pseudo-spin, and the effective mass anisotropy of each valley -- both the similarities and the differences in observed behaviors are significant.  
To facilitate such comparisons, it is useful to invoke the dimensionless parameter, $r_s \equiv 1/ ( a_B^\star\sqrt{\pi n}) $, where $n$ is the areal electron density, $a_B^\star$ is the effective Bohr radius.  $r_s= \bar V/\bar K$ is thus the ratio of a characteristic  electron-electron interaction strength, $\bar V$, to 
a characteristic kinetic energy, $\bar K$.

Because the 2DEG is burried deep in a heterostructure, many experiments that one would like to perform are not possible - the present studies depend entirely on measurements of the components of the resistivity tensor, $\rho_{ab}$. 
However, these have been  carried out with great precision as a function of the electron density, $n$ (controlled by a remote gate $150\mu$m from the 2DEG), applied magnetic field, both in-plane, $B_\parallel$, and out-of-plane, $B_\perp$,  shear strain to controlably break the underlying $C_4$ symmetry, and 
temperature, $T$. 
In an era in which  local  and/or time-resolved probes are  opening new horizons, it is worth recalling that most discoveries concerning the physics of quantum materials have stemmed from  measurements of the resistance.  
\begin{figure}[t]
\begin{center}
	\includegraphics[scale=0.5]{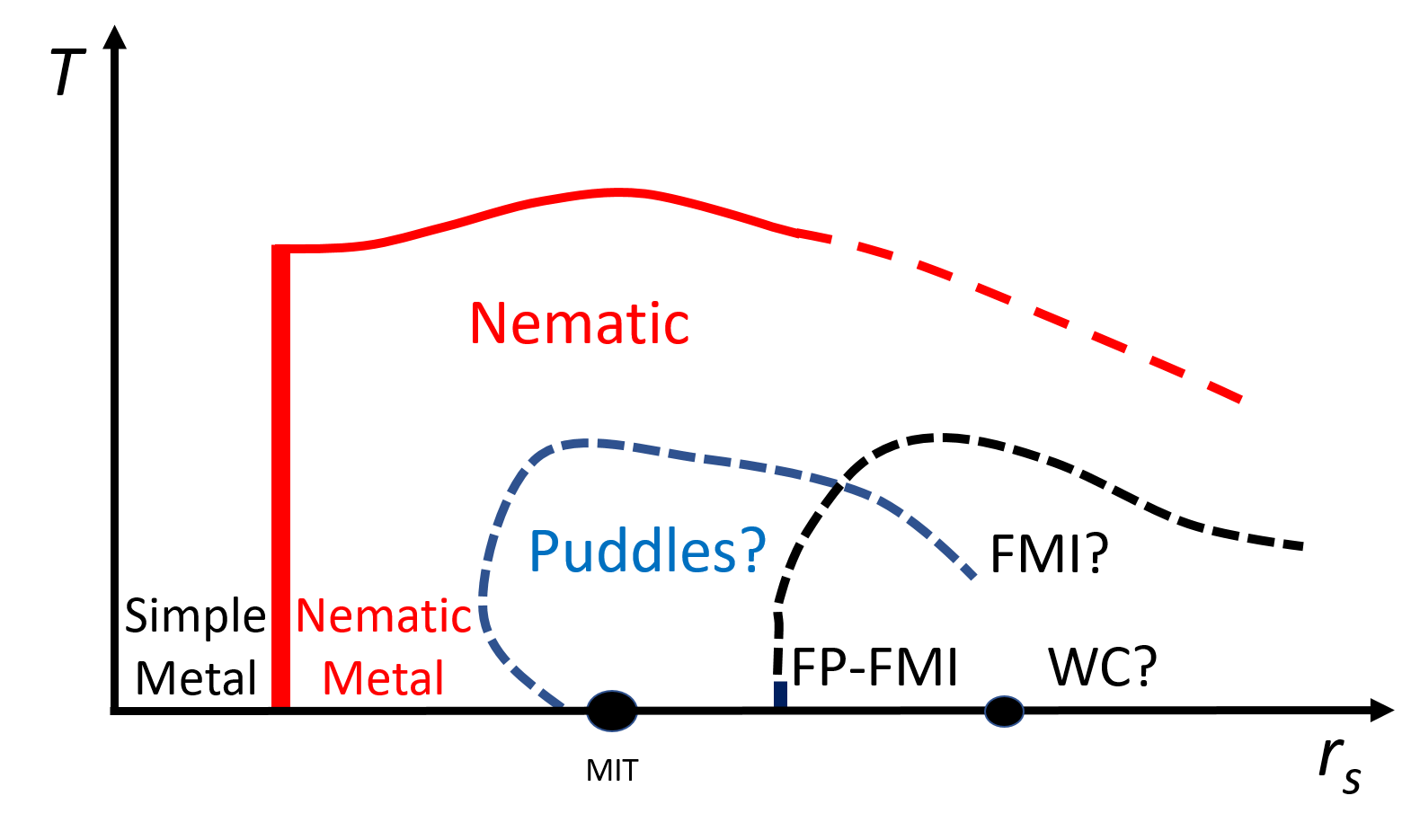}
\caption{Schematic phase digram of the 2DEG in AlAs.  Solid lines and circles represent transitions or sharp crossovers for which direct evidence is presented in Refs. \cite{Hossain2020ferromagnetism, hossain2020valley}.  The thick line represents a first order transition while the thin line is continuous.  The  dashed lines indicate boundaries suggested in the concluding theoretical discussion of the present paper.  Abbreviations are as follows:  MIT - metal-insulator transition;  FMI - Ferromagnetic insulator;  FP-FMI - fully polarized FMI;  WC - Wigner crystal. }
\label{Fig2}
\end{center}
\end{figure}

None-the-less, the absence of direct thermodynamic information means that much about the phase diagram has to be inferred from indirect arguments. 
Moreover, since the  primary focus of much of the study is on $T\to 0$ quantum phases of matter, there is an implicit assumption that no major changes in the physics occur at new emergent scales below the base temperature of $T=0.3$K.  With these caveats, we begin by summarizing the major inferences (See also the solid lines in the schematic phase diagram in Fig. 1.)
\begin{itemize}
\item{\it i)}  For $r_s < r_n \approx 20$, there is an isotropic (i.e. $C_4$ invariant) paramagnetic metallic phase.
\item{\it ii)}  There is a first order transition at $r_s =r_n\approx 20$ and $T\to 0$ to a fully valley polarized  metallic phase. 
Since this phase  spontaneously breaks the $C_4$ symmetry to $C_2$, it is an Ising nematic phase.  For $r_s > r_n$, as a function of increasing $T$, there is a finite $T$ continuous transition 
to the $C_4$-symmetric phase at $T_n(r_s)\approx 1.2$K for $r_s > r_n$.  
\item{\it iii)} At $r_s = r_{\textrm{mit}} \approx 27$ there is an apparent 
MIT. There has been considerable ``philosophical'' debate about what this means, given that a precise definition of a MIT necessarily involves an extrapolation to $T=0$.  However, from a practical ``physics'' perspective,  there is nothing subtle about this ``transition''  -- for $r_s < r_{\textrm{mit}}$, the resistivity is well below the quantum of resistance, $\rho_q=e^2/h$, and decreases strongly with decreasing $T$, while for $r_s > r_{\textrm{mit}}$, $\rho > \rho_q$ and is a strongly increasing function of decreasing $T$.  This is very similar to what is seen in a variety of other semiconductor heterostructures \cite{spivak2010colloquium}.
\item{\it iv)} For $r_s > r_F \approx 35$, the ground-state is a fully polarized ferromagnetic insulator (FP-FMI).  
The evidence of this (which we find compelling) is that the value of $B_{\parallel}$ necessary to achieve full polarization 
(i.e. beyond which $\rho_{xx}$ is $B_\parallel$ independent)  tends to zero as $r_s \to r_F^-$, while for $r_s > r_F$, $\rho_{xx}$ is essentially independent of $B_{\parallel}$. 
(It seems to us 
that it is an interesting open question whether or not there 
exists 
a range of $r_F^\star < r_s< r_F $ in which the 2DEG ground state is partially spin polarized.) 

\item{\it v)} A final change in behavior was observed at $r_{\textrm{wc}} = 38$;  for $r_s>r_{\textrm{wc}}$ the $I-V$ curves show pronounced non-linearities, behavior that the authors of Ref. \cite{Hossain2020ferromagnetism} associate with the existence of some form of moderately long-range Wigner-crystalline (WC)  order. 
While this is likely valid in some approximate sense, given that WC long-range order is not possible (in the presence of even weak quenched disorder) it is probably not possible to give a precise criterion for this crossover.  At any rate, also for $r_s > r_{\textrm{wc}}$ the 2DEG remains ferromagnetic and increasingly strongly insulating, the larger $r_s$.
\end{itemize}

Consistent with long-standing results of microscopic theory\cite{tanatar1989GS_2DEG,Louie1995QMC} and with decades of work on various realizations of the 2DEG, the simple metallic phase - presumably a Fermi liquid - is stable up to remarkably large values of $r_s$ in the present class of devices.  However, this gives way to various other phases at still larger $r_s$.  Two 
features 
of this evolution that are newly established are the existence of a fully orbitally polarized nematic metal \cite{hossain2020valley} and of a fully polarized ferromagnetic insulator \cite{Hossain2020ferromagnetism}. 
Indeed, it seems hard to escape the conclusion that at the largest accessible values of $r_s$, the ground-state is a ferromagnetic WC, which is presumably still 
nematic as well.

 Where $r_s$ is large, the interaction energy is the largest energy in the problem, meaning that there is no formal or intuitive justification for applying essentially perturbative methods, such as Hartree-Fock, random-phase approximation (RPA), or indeed any diagramatic approach to the theoretical analysis of this problem.  At large enough $r_s$, the problem is amenable to systematic strong-coupling  analysis\cite{wigner1934interaction,chakravarty1999wigner}, but strictly speaking this approach can only be used to explore the behavior deep in the 
 WC phase. To obtain theoretical understanding of the phases that occur at large but finite $r_s$, one  must either rely on essentially variational microscopic approaches or on more phenomenological arguments.  
 In Fig. 1, we have attempted to combine results from Refs. \cite{Hossain2020ferromagnetism,hossain2020valley}  -- indicated as solid lines --  with some speculative additions  largely based on theoretical considerations -- as dashed lines. 
 
There are two distinct theoretical arguments that lead to the conclusion that first order transitions are forbidden in 2D.  
The first - based on Imry-Ma arguments - invokes the effects of quenched disorder.
The second - based on Coulomb-frustrated phase separation - is a special feature of Coulomb interactions in 2D \cite{spivak2006ME}.  Both these arguments imply that where a first order transition is expected, instead there should occur a range of densities in which some form of ``puddle'' phase  arises - a mesoscopic version of phase separation in which regions of the sample are 
in an approximate sense in one of these phases and other regions are in the other.  
Despite this, empirically, the nematic  transition at $T=0$ appears to be first order;  this can be rationalized as it occurs where the system is highly conducting on both sides of the transition, which leads to strong screening both of any quenched disorder and of the long-range Coulomb interactions, likely meaning that any such bubble phase   occurs in an unobservably narrow range of $r_s$. 

These considerations do not apply to the transition to a WC, given that the WC is an insulating phase.  
Indeed, one of us and Spivak\cite{spivak2006ME} have argued that the physics generally associated with the MIT in 2DEGs is a reflection of the existence of micro-emulsion phases consisting of regions of insulating WC and regions of metallic liquid. 
This is consistent with the recent evidence \cite{sarachik2019MIT} that the MIT in Si MOSFETs is more of a percolation phenomenon than a true quantum phase transition. 
However, it is difficult to distinguish this intrinsic physics from the alternative disorder driven picture, in which 
the coexisting regions of WC and Fermi liquid reflect subtle differences in the local distribution of impurities or other structural defects \cite{sarma2005two}. 
\footnote{Concerning the role of disorder in the MIT:   The fact that the MIT is  observed in the cleanest achievable 2DEGs and that the phenomena look similar in such different platforms as Si MOSFETS and modulation-doped AlAs quantum wells, argues that there is likely a large intrinsic character to any puddle formation - even if at the end of the day disorder proves to be important in pinning the puddles.} 
We thus speculate the existence of a ''bubble'' regime in the phase diagram without specifying the degree to which disorder is the driving force.  The MIT 
occurs 
when the  liquid portions cease to percolate.  
The backward slope shown for  the left edge of the bubble regime is reminiscent of the Pomeranchuk effect in He -- it reflects the fact that the low energy scale associated with exchange interactions in the WC implies that it is generally a higher entropy phase than the liquid\cite{spivak2006ME}.

There is one other striking observation in \cite{Hossain2020ferromagnetism} that can be interpreted as the evidence of the existence of such a bubble state at large $r_s$, i.e. deep in the insulating regime. 
When a perpendicular magnetic field, $B_\perp$, is applied to the system with $r_s > r_{\textrm{mit}}$, the longitudinal resistance at first increases strongly, but then shows pronounced minima at fields corresponding to a full Landau level, $\nu=1$, and a partially filled Landau level, $\nu = 1/3$.  Moreover, $\rho_{xy}$ exhibits a plateau at the same fields with values $\rho_{xy} \approx (h/e^2)$ and $\rho_{xy} \approx 3 (h/e^2)$ respectively.
However, this is not a quantum Hall liquid since
at $T=0.3$K and $r_s = 38$, $\rho_{xx}(\nu=1) \approx 30(h/e^2)$ and   $\rho_{xx}(\nu=1/3) \approx 75(h/e^2)$. 
In a quantum Hall liquid 
 $\rho_{xx}$ should vanish as $T\to 0$.   Put another way, all components of the conductivity tensor, $\sigma_{ab}$, are very far from their expected values in a quantum Hall state.  
This behavior is an approximation of a ``quantized Hall insulator,''\cite{shahar1997nature}.\footnote{It is mentioned in Ref. \cite{Hossain2020ferromagnetism} that $\rho_{xx}$ is a weakly decreasing function of decreasing $T$ in the regime we have identified as a quantum Hall insulator;  however, in the observable range of $T$, $\rho_{xx}$ is one to two orders of magnitude larger than the quantum of resistance, and the $T$ dependence is relatively weak.}
It is the behavior expected from a macroscopic mixture of small puddles of a quantum Hall liquid in an insulating background \cite{dykhne1994theory}.\footnote{Note that even for $r_s > r_{\textrm{wc}}$, where one might think that the system is a uniform (pinned) WC at $B_\perp = 0$, puddles of quantum Hall liquids might still arise at large $B_\perp$ since, as shown in Ref \cite{Jain2018crystallization}, a quantum Hall liquid will typically compete more successfully with the WC than does the Fermi liquid.}

 Finally, we comment a bit on the nature of the ferromagnetism.  
 It was shown in Ref. \cite{chakravarty1999wigner} that at asymptotically large $r_s$, the localized spins in the WC of an isotropic 2DEG form a ferromagnetic state.  The result is delicate -- it depends on small differences between two- and three-particle exchange contributions -- and so could be changed by all sorts of microscopic considerations.\footnote{Taken at face value, the exchange couplings computed in Ref. \cite{chakravarty1999wigner} would be smaller than the measurement temperatures.  However, microscopic details, such as the thickness of the 2DEG and the  mass anisotropy, could change these results both qualitatively and quantitatively.}
 The energy scales involved at large $r_s$ are, moreover, exponentially small. 
 Still, in the present context, it is tempting to view the ferromagnetism as being a feature of the WC rather than of the metallic liquid.  This interpretation is consistent with the fact that the fully polarized ferromagnetic phase seems to extend to the largest accessible values of $r_s$, and that it onsets only within the insulating phase;  $ r_F > r_{\textrm{mit}}$.  It would be interesting to further explore  the magnetic response of the system in the neighborhood of $r_{\textrm{mit}}$.  For instance, while the experimental evidence that full ferromagnetic polarization onsets at $r_F$, 
 if some sort of puddle state indeed occurs, it would be natural to expect an onset of some degree of ferromagnetism at $r_F^\star < r_F$.

\bibliographystyle{unsrt}
\bibliography{ref.bib}
 
\end{document}